\documentclass[aps,prd,twocolumn,preprintnumbers,nofootinbib,superscriptaddress,amsmath]{revtex4-2}
\usepackage[english]{babel}
\usepackage[pdftex, pdftitle={Article}, pdfauthor={Author}]{hyperref} 
\usepackage{dcolumn}
\usepackage{bm}
\usepackage{multirow}
\usepackage{comment} 
\usepackage[normalem]{ulem}

\usepackage{float}
\usepackage{amsmath}
\usepackage{amsfonts,amssymb}
\usepackage{hyperref}
\usepackage{graphicx}
\usepackage{natbib}
\usepackage[utf8]{inputenc}
\usepackage{xcolor}
\usepackage{bm}
\usepackage{diagbox}
\usepackage{hhline}
\usepackage[normalem]{ulem}

\usepackage[normalem]{ulem}

\usepackage{xcolor}
\definecolor{mypurple}{rgb}{0.50,0.00,0.50}
\definecolor{mygreen}{rgb}{0.00,0.45,0.10}

\usepackage{textcomp}

\newcommand{\be}{\begin{equation}}
\newcommand{\ee}{\end{equation}}

\newcommand{\bea}{\begin{eqnarray}}
\newcommand{\eea}{\end{eqnarray}}
\newcommand{\bean}{\begin{eqnarray*}}
\newcommand{\eean}{\end{eqnarray*}}

\newcommand{\id}{{\rm 1\kern -2.5pt I}}

\begin{document}

\title{On the geometrical and dynamical distinction between Unimodular and General Relativistic wormholes}

\author{Marco Bosquez}
\email{marco.bosquez@usach.cl}
\affiliation{Instituto de F\'{\i}sica, Pontificia Universidad Cat{\'o}lica de Valpara\'{\i}so,
Avenida Universidad 330, Curauma, Valparaíso, Chile.}
\author{Erick Pastén}
\email{erick.contreras@usm.cl}
\affiliation{Departamento de F\'{i}sica, Universidad de Santiago de Chile,
Avenida V\'{i}ctor Jara 3493,  Estaci\'{o}n Central, 9170124, Santiago, Chile}
\author{Mauricio Cataldo}
\email{mcataldo@ubiobio.cl}
\affiliation{Departamento de F\'{i}sica, Universidad del Bío-Bío,
Avenida Collao 1202, Concepción, Chile}.
\author{Norman Cruz}
\email{norman.cruz@usach.cl}
\affiliation{Departamento de F\'{i}sica, Universidad de Santiago de Chile,
Avenida V\'{i}ctor Jara 3493,  Estaci\'{o}n Central, 9170124, Santiago, Chile}
\affiliation{Center for Interdisciplinary Research in Astrophysics and Space Exploration (CIRAS), Universidad de Santiago de Chile, Avenida Libertador Bernardo O’Higgins 3363, Estación Central, 9170022 Santiago, Chile}


\begin{abstract}
We characterize traversable wormholes in Unimodular Gravity (UG) and investigate what distinguishes them from their General Relativistic (GR) counterparts. For a class of static and spherically symmetric solutions, we analyze their embedding and timelike geodesics, showing that the geodesic structure is entirely determined by the metric and is therefore identical in both theories.

To identify the physical origin of their differences, we compare the source sectors required to sustain the same wormhole geometry. We find that preserving a fixed UG geometry while imposing energy-momentum conservation requires restricting the equation of state, whereas relaxing this condition introduces an effective inhomogeneous vacuum contribution. We further show that the degree of exoticity is preserved even when energy-momentum is not conserved, while departures from the GR sector are encoded in an effective inhomogeneous vacuum structure whose asymptotic behavior resembles that of a cosmological constant.

Our results reinforce that UG gravity is geometrically equivalent to GR, while its distinctive features emerge in the dynamical interpretation of the source sector. More generally, our analysis illustrates that different gravitational theories may give rise to identical spacetime geometries while requiring different sources to sustain them .
\end{abstract}

\maketitle

\section{Introduction}

One of the central questions in gravitational physics concerns the relation between spacetime geometry and its underlying sources. In general relativity (GR), this relation is encoded in the Einstein field equations \cite{Einstein1915}, where the energy--momentum tensor is locally conserved. However, it is well known that different physical interpretations can be assigned to the same geometry by rearranging terms between the geometric and matter sectors. This ambiguity manifests itself in the interpretation of the cosmological constant either as a geometric term or as vacuum energy \cite{Weinberg1989}, as well as the use of effective fluids in modified gravity theories \cite{NojiriOdintsov2011,Sotiriou_2010}.

Unimodular gravity (UG) provides a particularly interesting framework in which this relation is modified at a fundamental level. By restricting the determinant of the metric, the theory reduces the diffeomorphism invariance of GR to volume-preserving transformations and leads to field equations corresponding to the traceless part of the Einstein equations \cite{Einstein1919,AndersonFinkelstein1971,Unruh1989,Alvarez2015}. A key consequence is that the cosmological constant emerges as an integration constant and not as a fundamental parameter introduced in the action. Moreover, in its general formulation, the covariant divergence of the energy--momentum tensor is not necessarily zero, allowing for an exchange of energy--momentum between matter and geometry \cite{Fabris_2022}. This feature challenges the standard interpretation of sources and raises the question of whether geometries that are equivalent at the metric level correspond to equivalent  configurations across different gravitational theories.

Traversable wormholes provide a natural setting to explore this issue. Originally introduced within GR as solutions connecting two asymptotically flat regions \cite{MorrisThorne1988,Visser1989}, these geometries are known to require violations of the null energy condition (NEC), typically interpreted as the presence of exotic matter \cite{MorrisThorneYurtsever1988}. This requirement has motivated the study of wormholes in alternative gravitational frameworks, including unimodular gravity, where it has been suggested that such configurations might avoid the need for exotic matter \cite{Agrawal_2023}. However, recent analyses have shown that the need for exotic matter persists in unimodular wormholes, both from parametric studies \cite{CataldoCruz2025} and from more general covariant arguments based on the null Raychaudhuri equation \cite{PastenBosquezCruz2026}.

Beyond the question of energy conditions, these results suggest that the differences between UG and GR may be far more subtle than initially expected. This is consistent with previous work by \cite{Ellis_2011}, who demonstrated that the trace free formulation is functionally equivalent to Einstein's equations once an integrability condition is imposed, preserving vacuum solutions and standard junction conditions while promoting the cosmological constant to an integration constant. These results raise a more fundamental question: if UG and GR wormholes share the same spacetime geometry and local kinematical properties, what physical ingredients actually distinguish them?

In this work, we address this question by studying a class of static and spherically symmetric wormhole solutions in unimodular gravity and comparing them with their general-relativistic counterparts. We show that, while the geodesic structure is identical for a given metric, the associated matter content differs substantially between the two theories. In particular, we demonstrate that reproducing unimodular wormhole solutions within GR requires either imposing nontrivial constraints on the equation of state or introducing an effective inhomogeneous vacuum energy component. Using this framework, we characterize the admissible parameter space of barotropic unimodular wormholes, identify the subset corresponding to the general-relativistic limit, and analyze the effective vacuum structure generated by the non-conservation current. Our analysis highlights that the differences between unimodular gravity and general relativity are fundamentally dynamical rather than geometrical or kinematical.

More generally, our results reinforce that different gravitational theories may give rise to identical spacetime geometries while requiring different source sector to sustain them.

\section{Wormholes in unimodular gravity}
\label{sec:UG} 

Wormholes are defined in the framework of General Relativity as geometries connecting two asymptotically flat regions of spacetime \cite{MorrisThorne1988, Visser1989}. A well-known feature of these solutions is that they require violations of the null energy condition (NEC), typically interpreted as the presence of exotic matter \cite{MorrisThorneYurtsever1988}. This has motivated the exploration of alternative gravitational frameworks in which such geometries might be supported under different physical conditions.

Unimodular gravity (UG), originally proposed by Einstein \cite{Einstein1919}, modifies the dynamical structure of general relativity by restricting the metric determinant and effectively decoupling the cosmological constant from the gravitational action. A key feature of UG is that the metric determinant is fixed,
\begin{equation}
    |g_{\mu\nu}| = 1,
\end{equation}
which reduces the full diffeomorphism invariance of GR to the subgroup of volume-preserving diffeomorphisms. As a consequence, the field equations correspond to the traceless part of the Einstein equations,
\begin{eqnarray}
R_{\mu\nu} - \tfrac{1}{4} g_{\mu\nu} R 
= 8\pi G \left( T_{\mu\nu} - \tfrac{1}{4} g_{\mu\nu} T \right).\label{UGEquation}
\end{eqnarray}
In contrast, the Einstein equations in standard GR take the form
\begin{eqnarray}
R_{\mu\nu} - \tfrac{1}{2} g_{\mu\nu} R + \Lambda g_{\mu\nu} 
= 8\pi G T_{\mu\nu}.\label{GREquation}
\end{eqnarray}

In this context, wormhole solutions have been recently studied, with claims that certain configurations may avoid the need for exotic matter \cite{Agrawal_2023}. However, this conclusion has been revisited and corrected in subsequent works. In particular, a parametric analysis shows that the allowed wormhole configurations necessarily violate the energy conditions \cite{CataldoCruz2025}, while a covariant argument based on the null Raychaudhuri equation demonstrates that this violation is unavoidable in unimodular gravity \cite{PastenBosquezCruz2026}.

To make these considerations explicit, we focus on a class of Morris--Thorne wormholes described by the static, spherically symmetric line element:
\begin{equation}\label{eqn:metric}
ds^2 = dt^2 
- \frac{dr^2}{1 - \frac{b(r)}{r}} 
- r^2 \left( d\theta^2 + \sin^2\theta\, d\phi^2 \right).
\end{equation}

The shape function takes the form:
\begin{equation}
    \frac{b(r)}{r}= \left( \dfrac{r}{r_0} \right)^{\frac{2(\alpha + 1)}{1 + \alpha(2\beta - 1)}} =\left( \dfrac{r}{r_0} \right)^{\xi}\label{bdefinition}
\end{equation}
where $\alpha$ and $\beta$ parametrize a barotropic equation of state,
\begin{eqnarray}
    p_r=\alpha\rho, \\
    p_t=\beta p_r, \label{eq:EoS_t}
\end{eqnarray}
for the radial and tangential pressures, respectively. Using $\kappa=8\pi G$, the corresponding energy density and pressures are
\begin{eqnarray}
    \kappa\rho(r)&=&\frac{2C\,r^{-{\frac{4\alpha(\beta-1)}{1+(2\beta-1)\alpha}}}}{1+\alpha(2\beta-1)}, \\
    \kappa p_r(r)&=&\alpha \frac{2C\,r^{-{\frac{4\alpha(\beta-1)}{1+(2\beta-1)\alpha}}}}{1+\alpha(2\beta-1)},\\
    \kappa p_t(r)&=& \alpha\beta \frac{2C\,r^{-{\frac{4\alpha(\beta-1)}{1+(2\beta-1)\alpha}}}}{1+\alpha(2\beta-1)}. \label{EoS}
\end{eqnarray}

Here, $C$ is a constant fixed by the condition that the throat is located at $r=r_0$,
\begin{eqnarray}
    C=r_0^{\frac{-2(\alpha+1)}{1+\alpha(2\beta-1)}}. \label{Eq: C}
\end{eqnarray}
The flaring-out condition at the throat implies that the parameters $\alpha$ and $\beta$ must satisfy \cite{CataldoCruz2025}
\begin{equation}
    \frac{(\alpha+1)}{1+\alpha(2\beta-1)}  <0. \label{eqn:inequality}
\end{equation}

\begin{figure*}[t]
    \centering
    \begin{minipage}{0.74\textwidth}
        \raggedright
        \includegraphics[width=0.946\linewidth]{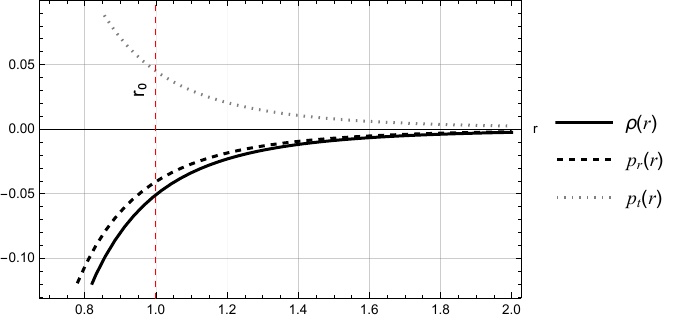}\\[0.15cm]
        \includegraphics[width=0.995\linewidth]{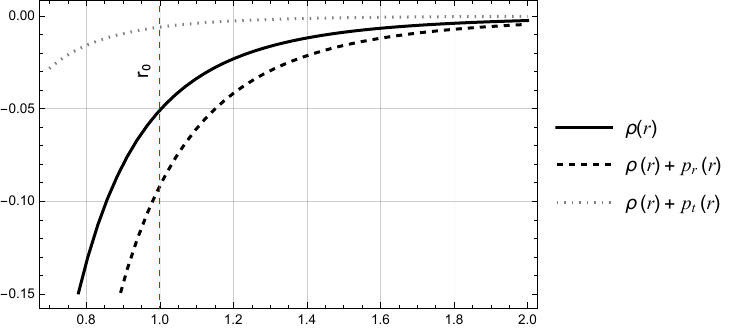}
    \end{minipage}

    \caption{
    Energy density $\rho(r)$ and combinations entering the null and weak
    energy conditions for representative values of the parameters. The vertical red dashed line indicates the wormhole
    throat at $r=r_0$. Both $\rho(r)$ and $\rho(r)+p(r)$
    remain negative throughout the spacetime, explicitly showing that the
    null and weak energy conditions are violated at all radii.
    }
    \label{EnergyConditions}
\end{figure*}

This constraint restricts the allowed parameter space to the following regions:
\begin{eqnarray}
    \alpha  <-1, \>\beta<\frac{\alpha-1}{2\alpha},  \\
    -1<\alpha<0, \>\frac{\alpha-1}{2\alpha}<\beta, \\
    0<\alpha, \>\beta<\frac{\alpha-1}{2\alpha}.
\end{eqnarray}

Within this domain, the combination entering the NEC becomes
\begin{equation}
    \kappa(\rho+p_r)=\frac{2C(\alpha+1)}{1+\alpha(2\beta-1)}\,r^{-{\frac{4\alpha(\beta-1)}{1+(2\beta-1)\alpha}}} <0,
\end{equation}
which is strictly negative for all $r$. Therefore, both the null and weak energy conditions are violated throughout the spacetime. This behavior is illustrated in Fig.~\ref{EnergyConditions}, where we plot $\rho(r)$ and the combinations $\rho+p_r$ and $\rho+p_t$ for representative values of the parameters.

These results confirm that UG does not remove the need for exotic matter in this class of wormhole solutions. More importantly, \cite{PastenBosquezCruz2026} suggest that the similarities between UG and GR extend beyond energy conditions. This observation is consistent with the trace-free formulation studied by \cite{Ellis_2011}, where vacuum solutions and standard junction conditions remain unchanged between UG and GR. This naturally motivates the question of whether this equivalence also extends to the geodesic structure of wormhole spacetimes. In the following section, we explicitly demonstrate that, for a given metric, UG and GR share the same geodesic structure.

\subsection{Geodesic structure}

Since the motion of test particles is entirely determined by the spacetime metric, any two gravitational theories sharing the same geometry must necessarily exhibit identical geodesic behavior. Therefore, geodesics provide an explicit and physically intuitive demonstration that the differences between UG and GR cannot be encoded in the kinematics of particles. Restricting the motion to the equatorial plane
$\theta=\pi/2$, the Lagrangian associated with the
metric~\eqref{eqn:metric} is

\begin{equation}
\mathcal{L}
=
\frac{1}{2}
g_{\mu\nu}
\frac{dx^\mu}{d\lambda}
\frac{dx^\nu}{d\lambda},
\end{equation}

where $\lambda$ is an affine parameter. The existence of
the timelike and axial Killing vectors leads to the conserved
quantities

\begin{align}
\dot t &= E,\\
r^2\dot\phi &= L,
\end{align}

which can be interpreted as the energy and angular momentum
per unit mass, respectively. Using these conserved quantities,
the radial equation of motion can be written as

\begin{equation}
\dot r^2
=
\left(
E^2-h-\frac{L^2}{r^2}
\right)
\left[
1-\left(\frac{r}{r_0}\right)^\xi
\right],
\label{eq:geodesic_summary}
\end{equation}

where $h=1$ for time-like geodesics and $h=0$ for null
geodesics. Equation~\eqref{eq:geodesic_summary} completely determines
the geodesic structure of the spacetime. The full derivation
of radial and non-radial trajectories, turning points,
traversability conditions, and orbital solutions for different barotropic parameters is presented
in Appendix~\ref{AppA}. Here we summarize only the main physical
results.

The geodesic analysis reveals two qualitatively distinct
classes of trajectories. Depending on the conserved quantities
$(E,L)$, particles either traverse the wormhole throat and
connect the two asymptotic regions or are reflected at a
finite-radius reversal point. The transition between both
regimes is determined entirely by the geometry and the
conserved quantities, confirming the traversable nature of
the wormhole spacetime.

\begin{figure*}
\centering
\includegraphics[width=0.8\textwidth]{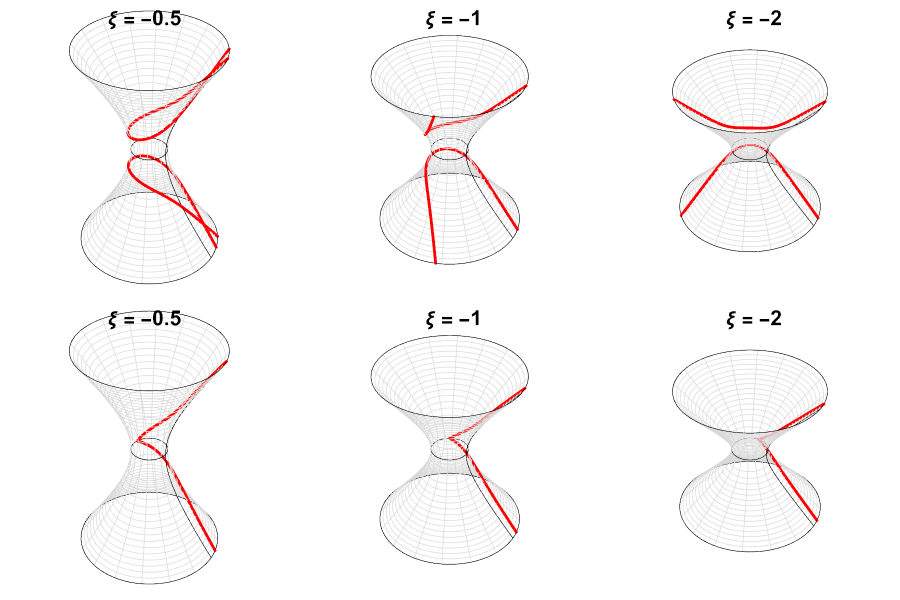}
\caption{
Representative timelike particle trajectories on the embedded
wormhole surfaces for $r_0=1$ and $L=6$. The columns correspond to
different values of the parameter $\xi=-0.5,-1,$ and $-2$,
while the upper and lower rows correspond to $E=6$ and $E=8$,
respectively. The red curves represent the particle trajectories.
For $E=6$, the particle is reflected before reaching the throat,
whereas for $E=8$ it has sufficient energy to cross the throat and
reach the opposite asymptotic region.}
\label{AllOrbits}
\end{figure*}

Figure~\ref{AllOrbits} illustrates representative
time-like geodesics embedded in the wormhole geometry.
Trajectories with sufficiently high energy cross the throat
and emerge in the opposite asymptotic region, while lower
energy trajectories are reflected before reaching the throat.
These behaviors arise solely from the metric structure and
are therefore identical in UG and GR. This complete geodesic analysis naturally shifts the focus from geometry to dynamics, motivating the source sector comparison developed in the following sections.

\section{Unimodular versus Einstein descriptions: source structure and effective dynamics}
\label{sec:UGvsGR}

To understand the physical distinction between UG wormholes and their general relativistic counterparts, it is necessary to analyze how each theory constrains the relation between sources. Although equations \ref{UGEquation} and \ref{GREquation} can be formally related by introducing an effective cosmological term, the underlying dynamical structure differs. In UG, one can define
\begin{eqnarray}
\Lambda_{\text{int}}(x) \;=\; \tfrac{1}{4}\,\big(R + 8\pi G\,T\big),
\end{eqnarray}
which acts as a spacetime dependent effective vacuum contribution.

A crucial difference emerges at the level of the energy--momentum conservation. In UG, the covariant divergence of $T_{\mu\nu}$ is not generically zero, but instead satisfies
\begin{equation}
    \nabla_\mu T^{\mu\nu} \equiv J^{\nu},
\end{equation}
where the current $J^\nu$ encodes a non-conventional exchange of energy--momentum between matter and geometry. Using the Bianchi identity, this current can be written as
\begin{eqnarray}
8\pi G\nabla_\mu \left(T^{\mu\nu} - \tfrac{1}{4} T g^{\mu\nu}\right)
= \tfrac{1}{4}\nabla_\mu (R g^{\mu\nu}),
\end{eqnarray}
leading to
\begin{equation}
    J^{\nu}
    = \frac{1}{8\pi G}\,\nabla_\mu \left( \tfrac{1}{4}R g^{\mu\nu} \right)
      + \nabla_\mu \left( \tfrac{1}{4} T g^{\mu\nu} \right).
\end{equation}

This non-conservation can be directly related to the effective vacuum term through
\begin{eqnarray}
\nabla_\mu \big(g^{\mu\nu} \Lambda_{\text{int}}\big) 
= \kappa J^\nu. \label{JLambda}
\end{eqnarray}

This relation highlights the central difficulty in embedding unimodular solutions into GR: a spacetime with a given geometry may correspond to inequivalent source configurations depending on whether one imposes conservation of $T_{\mu\nu}$ or allows for an effective vacuum contribution. In what follows, we explore different ways of comparing the UG wormhole solution with GR, making explicit the assumptions required in each case. We will call each comparison strategy as \textit{Strategy A, Strategy B and Strategy C}, respectively.

\subsection{Imposing energy conservation $\nabla_\mu T^{\mu\nu}=0$}\label{comparison 1}

A direct way to relate the unimodular solution to general relativity is to impose local conservation of the energy--momentum tensor, $\nabla_\mu T^{\mu\nu}=0$, as an additional equation.  In the unimodular framework, however, the current $J^\nu$ is generically nonzero. By symmetry, it can only have a radial component,
\begin{equation}
    J^{\nu} = (0, J^{r}(r), 0, 0).
\end{equation}

For the present geometry, one finds
\begin{eqnarray}
J^r &=& \nabla_\mu T^{\mu r} \\
&=& \left(1 - \frac{b(r)}{r}\right) 
\left[ \alpha \, \rho'(r) + \frac{2}{r} \alpha(1 - \beta) \, \rho(r) \right], \label{current}
\end{eqnarray}
where $\rho(r)$ is the energy density and $b(r)$ the shape function. Note that the anisotropy term is $(2/r)(p_r-p_t)$ with $p_r-p_t=\alpha(1-\beta)\rho$, consistently with the convention $p_t=\beta p_r$ of Eq.~\eqref{eq:EoS_t}. Imposing $J^\nu=0$ is equivalent, through Eq.~\eqref{JLambda}, to requiring that the effective vacuum contribution $\Lambda_{\text{int}}$ be constant. This has been extensively done in the literature to change the origin of the cosmological constant problem from an additional parameter of the theory to an integration constant (see e.g.\cite{Einstein1919, Weinberg1989}). However, this condition does not hold generically for the wormhole unimodular solution, and enforcing it introduces nontrivial constraints on the equation of state. Recalling that:

\begin{eqnarray}
    \xi=\frac{2(\alpha + 1)}{1 + \alpha(2\beta - 1)},\label{Eq..xi}
\end{eqnarray}

the constant $C$ in Eq.~\eqref{Eq: C} can be expressed as $C = r_0^{-\xi}$, which allows the energy density and pressures to be written in the compact form
\begin{eqnarray}
\kappa\rho(r)&=& \frac{\xi}{\alpha+1}r_0^{-\xi}r^{\xi-2}, \label{EoS_compact} \\
\kappa p_r(r)&=& \frac{\alpha\xi}{\alpha+1}r_0^{-\xi}r^{\xi-2},\\
\kappa p_t(r)&=& \frac{\alpha\beta\xi}{\alpha+1}r_0^{-\xi}r^{\xi-2}.
\end{eqnarray}
Using Eq.~\eqref{EoS_compact}, the current \eqref{current} reduces to
\begin{eqnarray}
J^r &=&
\frac{\alpha\,\xi\left(\xi - 2\beta\right)}{\kappa(\alpha+1)}
r_0^{-\xi}
\left[
r^{\xi-3} - r_0^{-\xi}r^{2\xi-3}
\right].
\label{eq:Jr_new}
\end{eqnarray}
Expression~\eqref{eq:Jr_new} vanishes identically if
\begin{eqnarray}
\beta = \frac{\xi}{2}.
\label{eq:GRcond}
\end{eqnarray}

Substituting $\beta$ from the condition~\eqref{eq:GRcond} into Eq.~\eqref{Eq..xi}, which defines $\xi$, we obtain
\begin{eqnarray}
\alpha\,\xi^2 + (1-\alpha)\xi - 2(\alpha+1) = \left(\xi-2\right)\left(\alpha\xi+\alpha+1\right)=0,\nonumber \\
\label{eq:xiquad}
\end{eqnarray}
whose admissible solution, satisfying $\xi < 0$, is given by
\begin{eqnarray}
{\xi_A(\alpha) =
-\frac{1+\alpha}{\alpha},
\qquad
\beta_A(\alpha)=\frac{\xi_A}{2}=-\frac{1+\alpha}{2\alpha}.}
\label{eq:xiGR}
\end{eqnarray}

This result shows that in this limit $\xi$ is no longer a free parameter but is fully determined by $\alpha$ alone, so that the  two parameter family $(\alpha,\beta)$ of UG solutions collapses to a one parameter family. A particularly clean exact case is $\alpha=1$, where Eq.~\eqref{eq:xiGR} gives $\xi=-2$ and $\beta=-1$ exactly. 

\subsection{$\Lambda_{\text{int}}$ as an effective exotic source}
\label{comparison2}

An alternative way to embed the unimodular solution into GR is to allow for a spacetime dependent effective vacuum contribution. In this approach, $\Lambda_{\text{int}}(x)$ is interpreted as part of the source sector rather than as a geometric term \cite{Fabris_2022}. We define the effective stress--energy tensor associated with $\Lambda_{\text{int}}(x)$ as
\begin{eqnarray}
T_{\mu\nu}^{(\Lambda)} \equiv - \frac{\Lambda_{\text{int}}(x)}{8 \pi G} \, g_{\mu\nu},
\end{eqnarray}
so that the total stress--energy tensor becomes
\begin{eqnarray}
T_{\mu\nu}^{\text{tot}} 
= T_{\mu\nu}^{(m)} + T_{\mu\nu}^{(\Lambda)}.
\end{eqnarray}

With this definition, the field equations take the standard form
\begin{eqnarray}
R_{\mu\nu} - \tfrac{1}{2} R g_{\mu\nu} 
= 8 \pi G \, T_{\mu\nu}^{\text{tot}}, \label{eq:EinsteinEffective}
\end{eqnarray}
with a conserved total energy--momentum tensor,
\begin{eqnarray}
    \nabla_\mu T^{\mu\nu}_{\text{tot}} = 0.
\end{eqnarray}

In this picture, the metric remains unchanged, but the physical interpretation of the source is modified. By symmetry, $\Lambda_{\text{int}}(x)=\Lambda_{\text{int}}(r)$, and the associated effective fluid can be written as
\begin{eqnarray}
\rho_\Lambda(r) = \frac{\Lambda_{\text{int}}(r)}{8\pi G}, 
\qquad
p_\Lambda(r) = -\rho_\Lambda(r).
\end{eqnarray}

The total energy density and pressures then become
\begin{eqnarray}
\rho_{\rm tot} &=& \rho + \rho_\Lambda, \\
p_{r}^{\rm tot} &=& \alpha \, \rho - \rho_\Lambda, \\
p_{t}^{\rm tot} &=& \alpha\beta \, \rho - \rho_\Lambda.
\end{eqnarray}

This leads to effective, radius-dependent equations of state,
\begin{eqnarray}
w_r^{\rm (eff)}(r) &=& \frac{\alpha \rho - \rho_\Lambda}{\rho + \rho_\Lambda}, \\
w_t^{\rm (eff)}(r) &=& {\alpha\beta \rho - \rho_\Lambda}{\rho + \rho_\Lambda}.
\end{eqnarray}

The function $\Lambda_{\text{int}}(r)$ is determined by the non-conservation current through Eq.~\eqref{JLambda},
\begin{eqnarray}
\Lambda_{\text{int}}'(r) 
= -8\pi G \left[ \alpha\,\rho'(r) + \frac{2}{r}\alpha(1-\beta)\,\rho(r) \right],
\end{eqnarray}
which can be integrated as
\begin{align}
\Lambda_{\text{int}}(r) 
= &\Lambda(r_0) \\
&- 8\pi G \int_{r_0}^{r} \left[ \alpha\, \rho'(s) 
+ \frac{2}{s}\alpha(1 - \beta)\, \rho(s) \right] ds.
\end{align}

Here $\Lambda(r_0)$ plays the role of a normalization constant. If we set $\Lambda(r_0)=0$ to isolate the radial variation induced by the non-conservation current, we have
\begin{eqnarray}
\rho_\Lambda(r) &=& -\int_{r_0}^{r} \left[ \alpha\, \rho'(s) 
+ \frac{2}{s}\alpha(1 - \beta)\, \rho(s) \right] ds.
\end{eqnarray}

\begin{figure*}[t] 
    \centering
    \begin{tabular}{cc} 
        \includegraphics[width=0.5\textwidth]{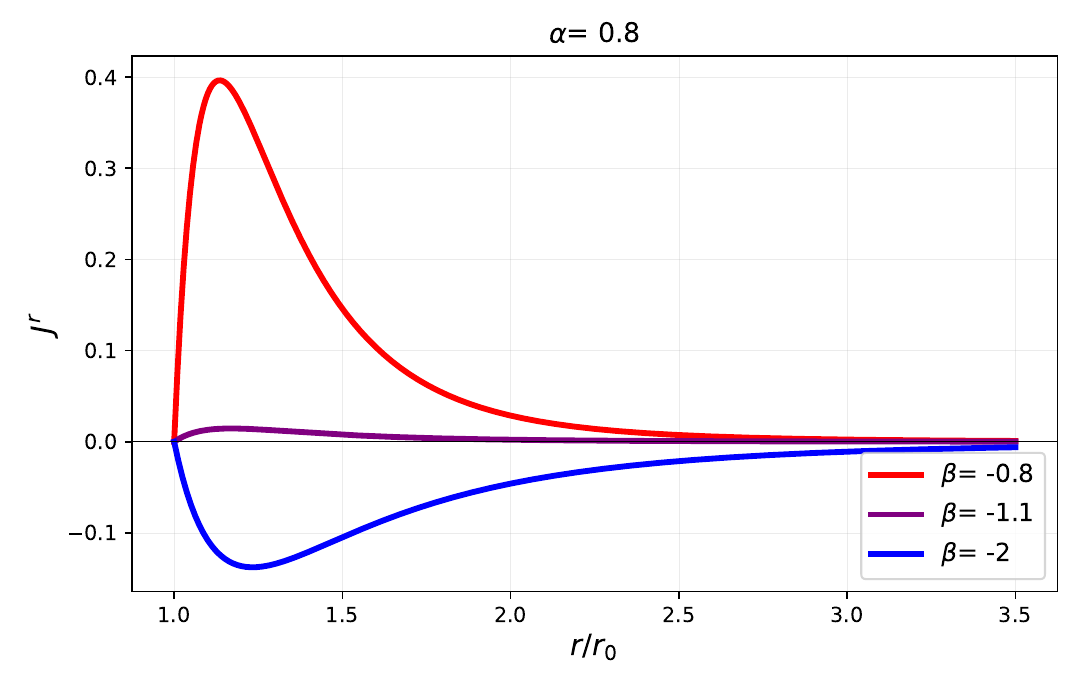} &
       \includegraphics[width=0.5\textwidth]{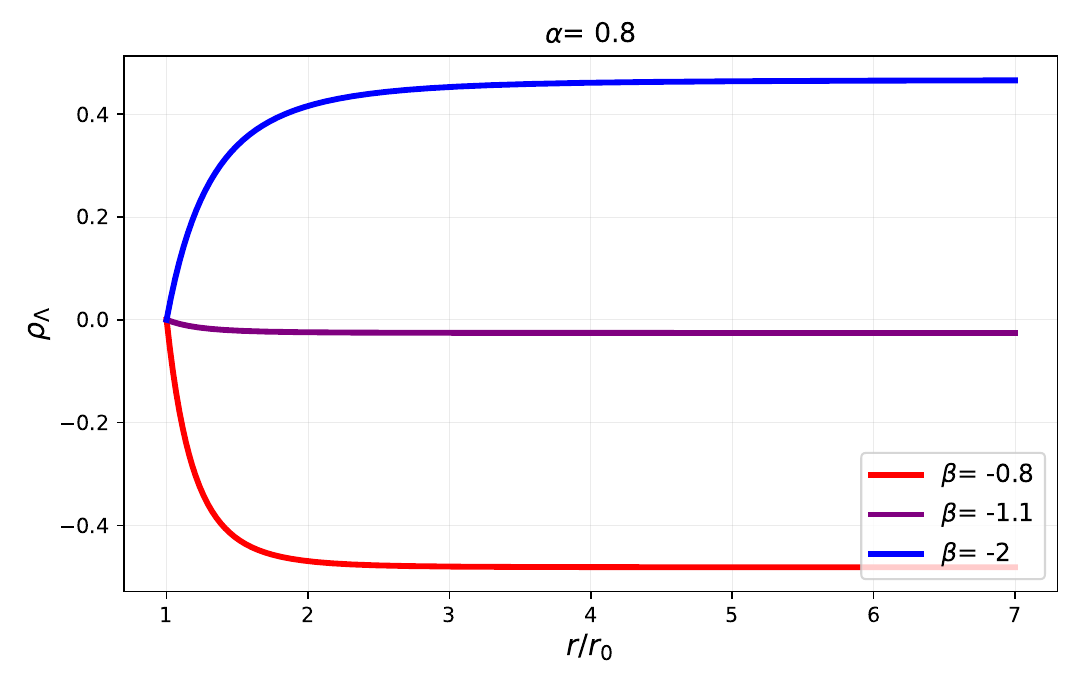} \\  
    \end{tabular}
    \caption{
Radial energy--momentum current $J^r(r)$ (left) and effective energy density
$\rho_\Lambda(r)$ (right) associated with the spacetime dependent
$\Lambda_{\mathrm{int}}(r)$, assuming $\Lambda(r_0)=0$.
The non-vanishing current reflects the exchange between matter and geometry
in unimodular gravity, while $\rho_\Lambda$ represents the effective
inhomogeneous vacuum contribution required in GR.}
    \label{fig:Jrrho}
\end{figure*}

This construction shows that the same wormhole geometry can be supported in GR by introducing an effective inhomogeneous vacuum energy component. However, this comes at the cost of replacing the original barotropic equation of state with a position dependent effective fluid, whose properties are entirely determined by $\Lambda_{\text{int}}(r)$. In particular, while the geodesic structure remains unchanged, the matter content acquires a nontrivial radial dependence, reflecting the underlying non-conservation present in the unimodular description (see Fig. \ref{fig:Jrrho}).

\subsection{The general relativistic limit as a consistency check}

Eqs.~\eqref{eq:xiGR} must coincide exactly with the unique solution obtained by solving the full GR Einstein equations for the metric~\eqref{eqn:metric} with $p_r=\alpha\rho$ and $p_t=\beta p_r$, as required by the Bianchi identity. To verify this explicitly, we now solve the Einstein equations directly under the same assumptions. In contrast to unimodular 
gravity, where the geometry and the equation of state can be specified independently, 
the GR field equations impose additional constraints that couple $b(r)$, $\alpha$, 
and $\beta$. From the radial component of the Einstein equations, using $p_r=\alpha\rho$, 
one obtains
\begin{eqnarray}
b'(r)= -\frac{b(r)}{\alpha r},
\end{eqnarray}
while the tangential component, with $p_t=\alpha\beta\rho$, leads to the relation
\begin{eqnarray}
2\alpha\beta + \alpha + 1=0,
\qquad\text{i.e.}\qquad
\beta=-\frac{1+\alpha}{2\alpha}.
\end{eqnarray}
As a result, the shape function is uniquely determined by
\begin{eqnarray}
\frac{b(r)}{r}
=
\left(
\frac{r}{r_0}
\right)^{-\frac{1+\alpha}{\alpha}},
\end{eqnarray}
which coincides exactly with the geometry obtained in Strategy~A, 
Eq.~\eqref{eq:xiGR}. This agreement is not accidental but obligatory as 
the traceless UG equations supplemented with local energy-momentum 
conservation are equivalent to the full Einstein equations according to the Bianchi identity. The unique GR wormhole compatible with the barotropic 
ansatz is therefore robust, independent of which route one follows to reach it.

\section{Exotic matter convergence and physical viability of UG wormholes}

Since the distinction between UG and GR has now been shown to reside in the source sector, an interesting question is whether this difference can also be quantified through standard measures of exoticity. To address this, note that the existence of a traversable wormhole geometry does not by itself guarantee the physical admissibility of the solution. Indeed, the flaring-out condition $\xi<0$ ensures that the throat exists geometrically, but it does not constrain the total amount of NEC violating matter required to support
the spacetime. A solution may therefore be mathematically consistent while containing an infinite amount of exotic matter.
To distinguish physically admissible wormholes from merely geometric
solutions, we can quantify the integrated NEC violation.
For this purpose we employ the Volume Integral Quantifier (VIQ)
\cite{Visser2003,Lobo2005}, which measures the total exotic matter content
supporting the wormhole and allows the parameter space to be divided into
regions of finite and divergent physical support,
\begin{eqnarray}
\mathcal{I} = \oint\!\left(\rho+p_r\right)dV
= 4\pi\int_{r_0}^{\infty}\!\left(\rho+p_r\right)r^2\,dr.
\label{eq:VIQ}
\end{eqnarray}
In UG
the integrand becomes
$(\rho+p_r)r^2 = (\xi/\kappa)\,r_0^{-\xi}\,r^{\xi}$.
The integral~\eqref{eq:VIQ} converges for $\xi<-1$, yielding 
\begin{eqnarray}
\mathcal{I}_{\mathrm{UG}} =
-\frac{4\pi}{\kappa}\,\frac{\xi(\alpha,\beta)}{\xi(\alpha,\beta)+1}\,r_0.
\label{eq:IVIQ_UG}
\end{eqnarray}
Since $\xi<-1$ implies $\xi/(\xi+1)>0$,
one has $\mathcal{I}_{\mathrm{UG}}<0$,
confirming the presence of exotic matter throughout.
The expression~\eqref{eq:IVIQ_UG} depends on the two free
parameters $(\alpha,\beta)$ through $\xi(\alpha,\beta)$.
This two parameter freedom is a direct consequence of the
non-conservation $J^\nu\neq 0$ in UG.

On the other hand, if one imposes energy conservation as in Strategy A the metric is unchanged, so $\rho$ and $p_r=\alpha\rho$
are identical to those of the UG solution.
Since the VIQ depends only on $\rho+p_r$, it follows that
\begin{eqnarray}
\mathcal{I}_{A}
= -\frac{4\pi}{\kappa}\,
\frac{\xi_A(\alpha)}{\xi_A(\alpha)+1}
\,r_0
\;=\;-\frac{4\pi}{\kappa}\,(1+\alpha)\,r_0,
\label{eq:IVIQ_GR}
\end{eqnarray}
where $\xi_A(\alpha)$ now depends on a single parameter and is given by Eq.~\eqref{eq:xiGR}. The last equality follows from $\xi_A/(\xi_A+1)=1+\alpha$. Since $\xi_A(\alpha)=-(1+\alpha)/\alpha<-1$ for all $\alpha>0$, the GR wormhole always carries a finite amount of exotic matter, growing linearly with $\alpha$.

\begin{figure*}[t]
\centering
\includegraphics[width=1\textwidth]{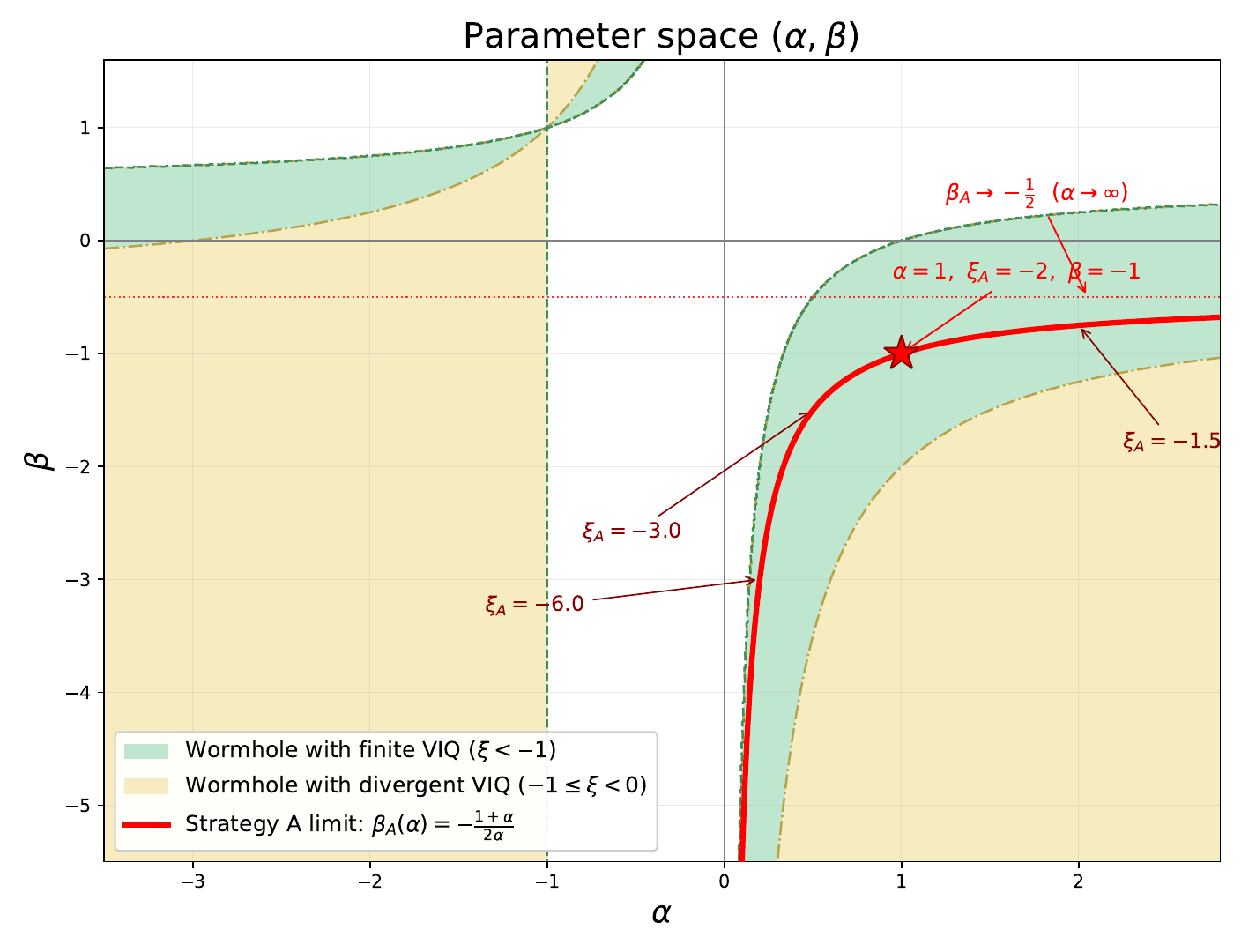}
\caption{
Structure of the $(\alpha,\beta)$ parameter space for the barotropic
unimodular wormhole solutions. The \textbf{white region} contains no
traversable wormhole, the \textbf{yellow region} supports geometric wormholes
but the VIQ diverges and the \textbf{green region} is the physically admissible
wormhole sector. The \textbf{red curve} is the Strategy A limit defined by
$\beta_A(\alpha)=-(1+\alpha)/(2\alpha)$ (Eq.~\eqref{eq:GRcond}), lying
entirely within the green region for all $\alpha>0$, with asymptote
$\beta_A\to -1/2$ as $\alpha$ approaches infinity. The red star marks the exact case
$\alpha=1$, $\xi_A=-2$, $\beta=-1$. Selected values of $\xi_A$ are indicated
along the curve.}
\label{fig:param_space}
\end{figure*}

Note that the VIQ formula~\eqref{eq:IVIQ_UG} involves only $\rho+p_r=(1+\alpha)\rho$,
which depends on the density and the radial equation of state. Therefore, it is insensitive to $p_t$ and to the conservation of $T_{\mu\nu}$. The VIQ itself cannot distinguish a UG solution from this GR limit, but the distinction is in the accessible range of $\mathcal{I}$. In UG one can tune $\beta$ to reach any exotic matter content at fixed $\alpha$,
whereas in this GR limit this freedom is locked by condition~\eqref{eq:GRcond}. The results derived above can be geometrically represented in the $(\alpha,\beta)$ parameter plane shown in
Fig.~\ref{fig:param_space}.
The plane is partitioned into three regions, each characterized by a different physical status of the barotropic solution.

The white region corresponds to parameter pairs for which the flaring-out condition fails ($\xi \geq 0$), so no wormhole throat exists and the solutions are not traversable. The yellow region satisfies the flaring-out condition ($-1 \leq \xi < 0$) but the VIQ integral diverges, indicating an infinite total amount of exotic matter.
The green region ($\xi < -1$) constitutes the physically
complete sector in which the throat exists and the
total exotic matter is finite. For any fixed $\alpha$, the UG complete family spans a continuous interval
of $\beta$ values (and hence a continuous range of $\xi$ and
exotic matter content), whereas the Strategy A limit selects a single point
$\beta = \beta_{A}(\alpha)$.
Displacing $\beta$ from the red curve activates the current $J^\nu$,
introduces a spatially varying $\Lambda_{\mathrm{int}}(r)$, and allows the wormhole to access a wider range of exotic matter configurations. 

The red curve, defined by condition~\eqref{eq:GRcond}, lies entirely
within the green region of finite exotic matter support.
The curve is the hyperbola $\beta_A(\alpha)=-(1+\alpha)/(2\alpha)$, which begins at $\alpha \to 0^+$, where $\beta_{A} \to -\infty$, and extends over the whole range $\alpha>0$, approaching the horizontal asymptote $\beta_A\to-1/2$ as $\alpha\to\infty$. Since $\xi_A(\alpha)=-(1+\alpha)/\alpha<-1$ for every $\alpha>0$, the curve never reaches the boundary $\xi=-1$ of the divergent-VIQ region.  Every $\alpha>0$ supports a GR wormhole with finite exotic matter content.

Finally, note that in the case of the spatially varying $\Lambda_{\rm int}(r)$ of Strategy B the geometry is identical to the UG solution, so the metric exponent
$\xi$ and the matter variables $\rho$ and $p_r=\alpha\rho$ are unchanged.
The total effective sources are
\begin{eqnarray}
\rho_{\mathrm{tot}} = \rho + \rho_\Lambda,
\qquad
p_r^{\mathrm{tot}} = \alpha\rho - \rho_\Lambda,
\end{eqnarray}
where $\rho_\Lambda=\Lambda_{\mathrm{int}}(r)/\kappa$ and
$p_\Lambda=-\rho_\Lambda$. Therefore, the $\Lambda$ contribution cancels
exactly from the NEC combination,
\begin{eqnarray}
\rho_{\mathrm{tot}} + p_r^{\mathrm{tot}}
= (\rho+\rho_\Lambda)+(\alpha\rho-\rho_\Lambda)
= (1+\alpha)\rho .
\end{eqnarray}
This means that the VIQ receives no direct contribution from
$\Lambda_{\mathrm{int}}(r)$ and is identical to the UG result:
\begin{eqnarray}
\mathcal{I}_B
=
\mathcal{I}_{\mathrm{UG}}
\label{eq:IB}
\end{eqnarray}
Therefore, Strategy~B changes the
interpretation of the source sector, but not the integrated NEC violating content measured by the VIQ.

\section{Asymptotic behavior of the effective vacuum sector}

In the previous sections we showed that imposing local energy--momentum conservation in UG reduces the original two parameter family of unimodular wormhole solutions to a one parameter, corresponding to the exact GR limit and coinciding with the solution of the full Einstein equations by virtue of the Bianchi identity. On the other hand, Strategy B preserves the full unimodular solution, and the non-conserved sector is reinterpreted as an effective vacuum contribution. This raises the following question: if the distinction between UG and GR resides in the source sector, how is this difference encoded in the effective vacuum structure generated by the non-conservation current?

As discussed in Sec.~\ref{comparison2}, the effective vacuum
sector is described by the function
$\Lambda_{\mathrm{int}}(r)$, whose radial evolution is directly
sourced by the non-conservation current $J^\nu$ through
\begin{eqnarray}
g^{rr}\Lambda_{\mathrm{int}}'(r)=\kappa J^r.
\label{eq}
\end{eqnarray}
Since $\Lambda_{\mathrm{int}}(r)$ acts as an effective vacuum
energy contribution, we characterize its asymptotic behavior by introducing the constant $\Lambda_{\mathrm{\infty}}$.
We therefore impose the boundary condition
\begin{eqnarray}
\lim_{r\to\infty}\Lambda_{\mathrm{int}}(r)
\equiv
\Lambda_{\mathrm{\infty}},
\label{eq:bc_infty}
\end{eqnarray}
where $\Lambda_{\infty}$ denotes the asymptotic constant value of the vacuum contribution. This condition is particularly natural in unimodular gravity,
where the cosmological constant emerges as an integration
constant rather than as a parameter of the action
\cite{Einstein1919,AndersonFinkelstein1971,Unruh1989,Alvarez2015}.

Using Eq.~\eqref{eq:Jr_new}, we obtain
\begin{eqnarray}
\Lambda_{\mathrm{int}}'(r)
=
-\frac{{\alpha\,\xi(\xi-2\beta)}}{\alpha+1}\,
r_0^{-\xi}r^{\xi-3}.
\label{eq:Lint_deriv}
\end{eqnarray}

Integrating Eq.~\eqref{eq:Lint_deriv} and imposing the boundary
condition~\eqref{eq:bc_infty}, we obtain the profile
\begin{equation}
\Lambda_{\mathrm{int}}(r)
=
\Lambda_{\mathrm{\infty}}
-
\frac{{\alpha\,\xi(\xi-2\beta)}}
{(\alpha+1)(\xi-2)}
\,r_0^{-\xi}r^{\xi-2}.
\label{eq:Lint_full}
\end{equation}

This expression shows that the effective vacuum energy generally differs from its asymptotic value at finite radii. To
quantify this deviation, we evaluate the solution at the wormhole
throat
\begin{eqnarray}
\Lambda_{\mathrm{int}}(r_0)
=
\Lambda_{\mathrm{\infty}}
+\Delta\Lambda,
\label{eq:Lint_throat}
\end{eqnarray}
where
\begin{eqnarray}
\Delta\Lambda
=
-\frac{{\alpha\,\xi(\xi-2\beta)}}
{(\alpha+1)(\xi-2)\,r_0^2}.
\label{eq:DLambda}
\end{eqnarray}

The quantity $\Delta\Lambda$ measures the difference between the effective vacuum energy at the throat and its asymptotic value.
Thus, although the constant $\Lambda_{\mathrm{\infty}}$ is recovered at spatial
infinity, the local vacuum energy near the throat is generally
shifted whenever $J^\nu\neq0$.

Remarkably, the condition
\begin{eqnarray}
\Delta\Lambda=0
\end{eqnarray}
is equivalent to
\begin{eqnarray}
{\xi-2\beta=0,}
\end{eqnarray}
which is precisely the condition defining the Strategy A discussed in
Sec.~\ref{comparison 1}. Note that the departure of UG from the GR sector can be quantified directly through the offset $\Delta\Lambda$.

\subsection{Physical interpretation and sign of $\Delta\Lambda$}

For $\xi < -1$ and $\alpha > 0$, the denominator of Eq.~\eqref{eq:DLambda} satisfies $(\alpha+1)(\xi-2)r_0^2 < 0$, since $(\alpha+1)>0$ and $(\xi-2)<0$. The numerator contains the factor $-\alpha\xi > 0$ (because $\alpha>0$ and $\xi<0$). Therefore the sign of $\Delta\Lambda$ is determined by the sign of $(\xi-2\beta)$ as follows:
\begin{itemize}
    \item If $\xi > 2\beta$: $\Delta\Lambda < 0$, so $\Lambda_{\mathrm{int}}(r_0) < \Lambda_\infty$. The local vacuum energy at the throat is below the asymptotic value.
    \item If $\xi < 2\beta$: $\Delta\Lambda > 0$, so $\Lambda_{\mathrm{int}}(r_0) > \Lambda_\infty$. The local vacuum energy at the throat exceeds the asymptotic value.
    \item If $\xi = 2\beta$: $\Delta\Lambda = 0$, so $\Lambda_{\mathrm{int}}(r_0) = \Lambda_\infty$. The vacuum energy is uniform throughout the spacetime. This is precisely the Strategy~A condition, i.e., the exact GR limit.
\end{itemize}
This last condition establishes a direct connection between the two main analyses of this paper. The Strategy~A condition $\beta = \xi/2$ is the unique curve in $(\alpha,\beta)$ space where the wormhole produces no gradient in $\Lambda_{\mathrm{int}}(r)$, so that the local and asymptotic values of the effective vacuum energy coincide. Off this curve, the wormhole generates a radial profile of $\Lambda_{\mathrm{int}}(r)$ controlled by the same factor $(\xi - 2\beta)$ that governs the energy--momentum current $J^\nu$.

The rate of approach to the asymptotic value is power-law,
\begin{eqnarray}
\Lambda_{\mathrm{int}}(r) - \Lambda_\infty \;\propto\; r^{\xi-2}, \qquad r\to\infty, \nonumber \\
\end{eqnarray}
with exponent $\xi - 2 < -3$ for $\xi < -1$, ensuring rapid convergence in the physically relevant range. Two limiting cases of $\Delta\Lambda$ are instructive. For a strongly flared geometry ($\xi \ll -1$),
\begin{eqnarray}
\Delta\Lambda \;\approx\; -\frac{\alpha\,\xi}{(\alpha+1)\,r_0^2},
\end{eqnarray}
which grows without bound and is independent of $\beta$. Near the VIQ convergence threshold ($\xi \to -1$),
\begin{eqnarray}
\Delta\Lambda \;\to\; \frac{\alpha(1+2\beta)}{3(\alpha+1)\,r_0^2}.
\end{eqnarray}

\subsection{Connection with the cosmological constant problem}

In GR, the cosmological constant is a fixed parameter of the action and cannot vary across the spacetime. In UG, by contrast, $\Lambda$ is a constant of integration that acquires a spatial profile $\Lambda_{\mathrm{int}}(r)$ whenever $J^\nu \neq 0$. The result~\eqref{eq:Lint_throat} shows that the wormhole geometry generates a local shift $\Delta\Lambda$ between the throat and the asymptotic region. The observed cosmological constant can be associated with the asymptotic value $\Lambda_\infty$, while the throat value $\Lambda_{\mathrm{int}}(r_0) = \Lambda_\infty + \Delta\Lambda$ can differ substantially depending on the wormhole parameters.

This mechanism is specific to UG and has no analogue in GR. In principle, it allows configurations where $\Delta\Lambda$ is large and negative, so that $\Lambda_{\mathrm{int}}(r_0) \ll \Lambda_\infty$, or large and positive, creating a region near the throat where the effective vacuum energy is much larger than its asymptotic value. We emphasize that this observation does not resolve the cosmological constant problem, but it identifies a new geometric mechanism within UG, in which the spatial variation of $\Lambda_{\mathrm{int}}(r)$ is sourced by the non-conservation current $J^\nu$. This mechanism is absent in GR and deserves further investigation. For example, whether an ensemble of such UG structures can contribute to an emergent cosmological background is an interesting question that can be addressed in future works.

\section{Conclusions}

In this work, we have shown that UG wormholes are geometrically indistinguishable from their GR counterparts. Since the motion of test particles is entirely determined by the spacetime metric, both theories exhibit the same geodesic structure whenever the geometry is fixed. The distinction between both theories therefore lies in the source sector.

This distinction reflects that energy-momentum conservation is not imposed in UG. If one imposes local conservation while keeping the same geometry, the original two-parameter barotropic family is reduced to a one-parameter subset. Relaxing this assumption preserves the full two-parameter family and introduces an effective inhomogeneous vacuum contribution $\Lambda_{\mathrm{int}}(r)$. Moreover, for a generic unimodular configuration with $\beta\neq\xi/2$, there is no map to GR that preserves the full source structure. Only on the curve $\beta=\xi/2$, the barotropic UG solution admits an exact GR interpretation.

The VIQ analysis provides a clean parametric description of unimodular wormholes, in which we identified regions where no traversable wormhole exists, regions where the geometry is admissible but the total exotic matter content diverges, and regions where the wormhole is physically complete. However, VIQ also reveals that the distinction between UG and GR cannot be reduced to the amount of exotic matter alone, since it is insensitive to source contributions that cancel in the NEC combination $\rho+p_r$. This means that different source sectors may share the same integrated exotic matter content, even when they differ in their conservation. We further show that, when energy conservation is not imposed, unimodular wormholes show a non-trivial vacuum structure, whose asymptotic behavior is reminiscent of a constant vacuum structure.

These results extend previous studies in Unimodular Gravity aimed at clarifying the physical distinctions between UG and GR. In GR, the cosmological constant may be interpreted either as an additional geometric parameter of the theory or as a constant vacuum-energy contribution. In UG, the cosmological constant can be interpreted as an integration constant when local energy--momentum conservation is imposed or, as shown here, as the asymptotic limit of an inhomogeneous vacuum structure generated by a compact object.

We conclude that wormhole space-times provide a useful laboratory to disentangle the kinematical and dynamical aspects of gravitational theories. Identical wormhole geometries can therefore correspond to inequivalent matter and vacuum sectors in UG and GR. This shows that spacetime geometry alone does not uniquely determine the physical content of a gravitational system. In this sense, wormhole space-times do not constitute suitable observational probes for discriminating between the two theories when they share the same geometry, since observables based on geodesic motion are insensitive to the differences encoded in the source sector.

\section*{Acknowledgements}

E. P. acknowledges support from the POSTDOC\_DICYT project 042531CM\_Postdoc, Vicerrectoría de Investigación, Innovación y Creación, Universidad de Santiago de Chile (USACH).

\appendix

\section{Geometric characterization and geodesic properties of UG wormholes}
\label{AppA}

\subsection{Embedding}

To gain geometric intuition, we consider spatial slices at constant $t$ and $\theta=\pi/2$, and embed them in a three-dimensional Euclidean space. This construction allows us to visualize the wormhole throat and the flaring-out condition in a coordinate-independent way.

Introducing $z(r)$ as the embedding function in cylindrical coordinates, the induced metric on the slice reads
\begin{equation}
    ds^2= \left(1+\left(\frac{dz}{dr}\right)^2\right)dr^2+r^2d\phi^2.
\end{equation}

Matching this expression with the spatial sector of the wormhole metric, we obtain
\begin{equation}\label{dz}
    \frac{dz}{dr}=\frac{1}{\sqrt{1-\left (  \frac{r}{r_0}\right )^{\xi}}},
\end{equation}
where
\begin{eqnarray}
    \xi=\frac{2(\alpha + 1)}{1 + \alpha(2\beta - 1)}<0. \label{Eq. xi}
\end{eqnarray}

Imposing the condition $z(r_0)=0$, the general solution is given by
\begin{equation}
z(r)=r\, {}_2F_1\left( \frac{1}{2}, \frac{1}{\xi}, 1 + \frac{1}{\xi}, \left( \frac{r}{r_0} \right)^{\xi} \right)
-\frac{\sqrt{\pi}\,\Gamma(1+\frac{1}{\xi})}{\Gamma(\frac{1}{2}+\frac{1}{\xi})},
\label{eqn:Embedding}
\end{equation}
where ${}_2F_1$ is the Gaussian hypergeometric function. For particular values of $\xi$, this expression simplifies. For instance, when $\xi=-1$, one finds
\begin{equation}
z=r\sqrt{1-\frac{r_0}{r}}+r_0\,\mathrm{arctanh}\left(\sqrt{1-\frac{r_0}{r}}\right).
\end{equation}

The scalar curvature of the embedded surface is
\begin{equation}
    \mathcal{R}(r,\xi)=\frac{(r/r_0)^\xi \, \xi}{r^2}.
\end{equation}
Since $\xi<0$, the curvature is negative everywhere, reflecting the flaring-out of the geometry at the throat.

\begin{figure}[h!]
\centering
\includegraphics[width=0.5\textwidth]{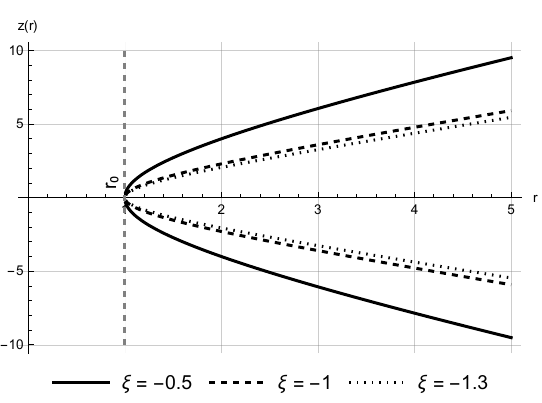}
\includegraphics[width=0.5\textwidth]{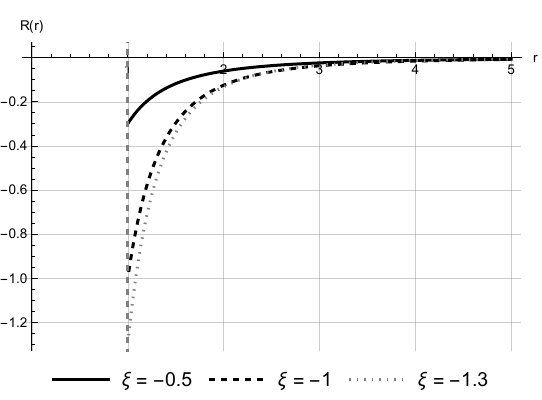}
\caption{
Embedding diagrams for different values of $\xi$. As $|\xi|$ increases, the curvature of the embedding surface becomes more negative, indicating a stronger flaring-out of the wormhole geometry near the throat.
}
\label{Embeddings}
\end{figure}

Embedding profiles for different values of $\xi$ are shown in Fig.~\ref{Embeddings}. As $|\xi|$ increases, the curvature becomes more negative, indicating a stronger flaring of the wormhole geometry.

\subsection{Geodesics}

To characterize the motion of test particles, we consider the geodesic Lagrangian
\begin{equation}
\mathcal{L}=\frac{1}{2}g_{\mu \nu} \frac{d x^\mu}{d \lambda}\frac{d x^\nu}{d \lambda},
\end{equation}
where $\lambda$ is an affine parameter. Restricting to the equatorial plane $\theta=\pi/2$, the Lagrangian associated with the metric \eqref{eqn:metric} becomes
\begin{equation}\label{eqn:lagrangian}
   h=\dot{t}^2-\frac{\dot{r}^2}{1 - \left( \dfrac{r}{r_0} \right)^\xi}-r^2\dot{\phi}^2,
\end{equation}
where $h=1$ for time-like geodesics and $h=0$ for null geodesics.

The existence of Killing vectors associated with time translation and axial symmetry leads to the conserved quantities
\begin{align}\label{eqn:cc}
    \dot{t}&=E, \\
    r^{2} \dot{\phi}&=L,
\end{align}
which can be interpreted as the energy and angular momentum per unit mass, respectively.

Substituting these expressions into \eqref{eqn:lagrangian}, the radial equation of motion can be written as
\begin{equation}\label{eqn:geodesic}
    \dot{r}^2 = \left(E^2-h-\frac{L^2}{r^2} \right)\left(1 - \left( \dfrac{r}{r_0} \right)^\xi\right).
\end{equation}

This equation fully determines the geodesic structure of the spacetime.

\subsection{Radial time-like geodesics}

We now analyze radial time-like geodesics, setting $L=0$ and $h=1$. In this case, Eq.~\eqref{eqn:geodesic} reduces to
\begin{equation}
    E^2-1 =\frac{\dot{r}^2}{1 - \left( \dfrac{r}{r_0} \right)^\xi}.
\end{equation}

Given initial conditions at $r=r_i$ with radial velocity $v_i$, the solution for the radial velocity is
\begin{equation}
    v(r)=\dot{r}=\pm v_{i}\sqrt{\frac{1-\left( \dfrac{r}{r_0} \right)^\xi}{1-\left( \dfrac{r_{i}}{r_0} \right)^\xi}}.
\end{equation}

The positive (negative) sign corresponds to outward (inward) motion. In the trivial case $v_i=0$, the particle remains at rest at $r=r_i$.

The proper time as a function of the radial coordinate can be written as
\begin{equation}
\tau(r)=\pm \frac{r}{v_i} \sqrt{1 - \left( \frac{r_i}{r_0} \right)^{\xi}} \,
{}_2F_1\left( \frac{1}{2}, \frac{1}{\xi}, 1 + \frac{1}{\xi}, \left( \frac{r}{r_0} \right)^{\xi} \right).
\end{equation}

\begin{figure}[h]
\centering
\includegraphics[width=0.5\textwidth]{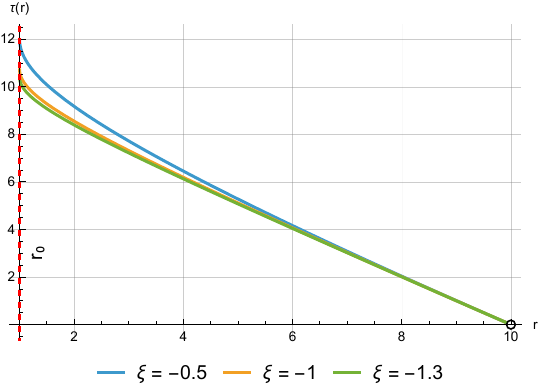}
\includegraphics[width=0.5\textwidth]{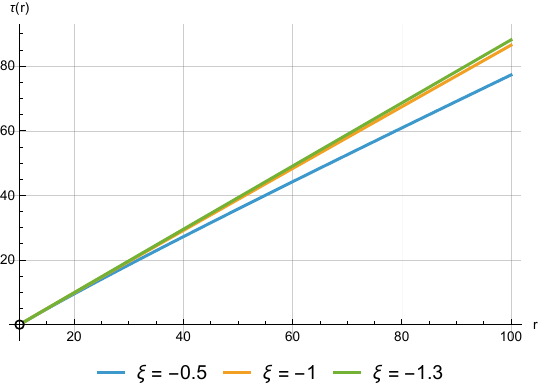}
\caption{
Proper time $\tau(r)$ for radial time-like geodesics with different values of $\xi$. The vertical red dashed line indicates the wormhole throat at $r=r_0$, while the white dot indicates the initial position $r=r_i$.}
\label{Tau}
\end{figure}

\begin{figure}[h]
\centering
\includegraphics[width=0.5\textwidth]{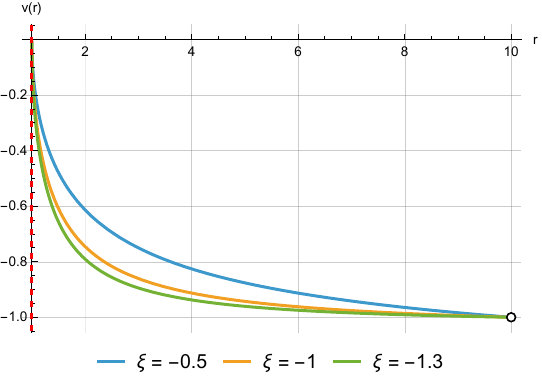}
\includegraphics[width=0.5\textwidth]{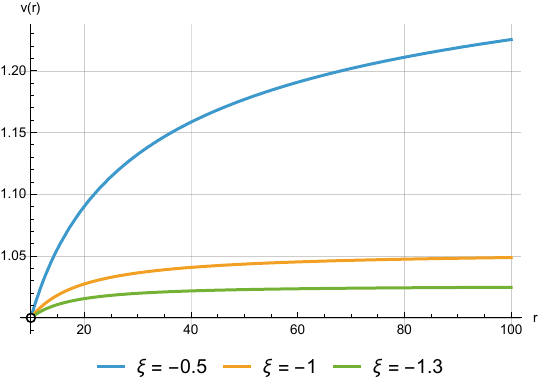}
\caption{
Radial velocity $v(r)$ for time-like geodesics. The vertical red dashed line indicates the wormhole throat at $r=r_0$. Particles with $v_i>0$ asymptotically approach a constant velocity as $r\to\infty$, while particles with $v_i<0$ fall toward the throat with vanishing velocity at $r=r_0$. Particles with $v_i=0$ remain at rest. 
}
\label{Vel}
\end{figure}

The behavior of $\tau(r)$ and $v(r)$ is shown in Figs.~\ref{Tau} and \ref{Vel}. Two qualitatively distinct regimes can be identified:

\begin{itemize}
\item \textbf{Inward motion ($v_i<0$):} particles approaching the wormhole throat always reach $r=r_0$ with vanishing radial velocity. This reflects the fact that $r$ attains a minimum at the throat.

\item \textbf{Outward motion ($v_i>0$):} particles moving away from the throat asymptotically approach a constant radial velocity as $r\rightarrow\infty$,
\begin{equation}
    v(r)\to v_i\left(1-\left(\frac{r_i}{r_0}\right)^\xi\right)^{-1/2}.
\end{equation}
\end{itemize}

These results show that the radial motion is entirely controlled by the geometric factor $1-(r/r_0)^\xi$. Note that the vanishing of $\dot{r}$ at the throat reflects the regularity of the geometry. As a consequence, this behavior is identical in UG and GR whenever the same metric is assumed.

\subsection{Non-radial time-like geodesics}

We now consider the case with non vanishing angular momentum, $L\neq 0$. For a massive particle, Eq.~\eqref{eqn:geodesic} evaluated at $r=r_i$ gives
\begin{equation}
    E^2=1+\frac{L^2}{r_i^2}+\frac{v_{i}^2}{1-\left(\frac{r_i}{r_0}\right)^{\xi}},
    \label{eqn:EnergyL}
\end{equation}
which relates the conserved energy to the initial conditions.

The radial velocity can then be written as
\begin{equation}
    \dot{r}^2=\left[L^2\left(\frac{1}{r_i^2}- \frac{1}{r^2}\right)+\frac{v_{i}^2}{1-\left(\frac{r_i}{r_0}\right)^{\xi}}\right]\left(1-\left(\frac{r}{r_0}\right)^{\xi}\right).
\end{equation}

This expression separates naturally into two contributions: a geometric term associated with the angular momentum $L$, and a term controlled by the initial radial velocity $v_i$ and the parameter $\xi$, which encodes the wormhole geometry. 

\subsubsection{Outward geodesics ($v_i>0,\ r>r_i$)}

\begin{figure}[h!]
\centering
\includegraphics[width=0.5\textwidth]{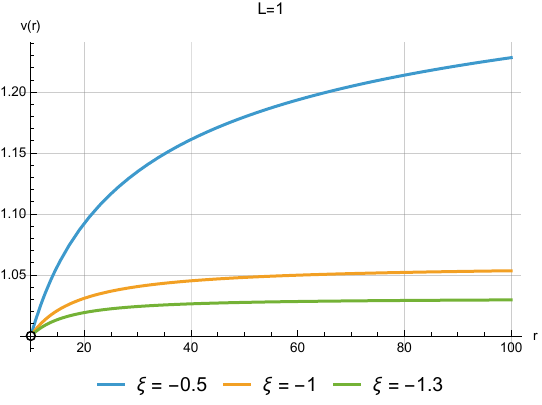}
\includegraphics[width=0.5\textwidth]{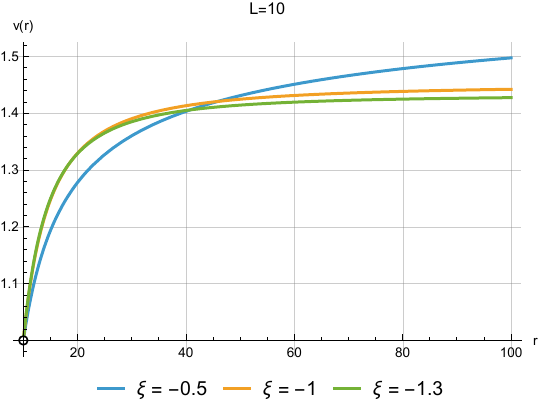}
\includegraphics[width=0.5\textwidth]{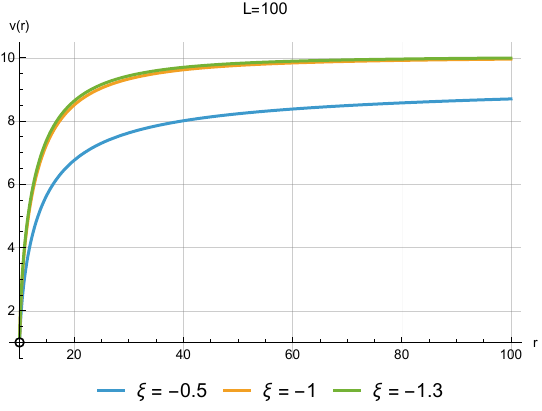}
\caption{
Radial velocity $v(r)$ for outward geodesics with different values of the angular momentum $L$. At large distances, the contribution from $L$ dominates the motion.
}
\label{VelL}
\end{figure}

For outward motion, $\dot{r}>0$ for all $r>r_i$. The behavior of $v(r)$ for different values of $L$ is shown in Fig.~\ref{VelL}. In the purely radial case ($L=0$), we found that more negative values of $\xi$ reduce the asymptotic velocity, reflecting the influence of the wormhole geometry.

When $L\neq 0$, however, the angular momentum term introduces an additional contribution that modifies this behavior and becomes dominant at large distances. In the limit $r\rightarrow\infty$, the radial velocity approaches
\begin{equation}
    \dot{r} \to \sqrt{\frac{L^2}{r_i^2}+\frac{v_{i}^2}{1-\left(\frac{r_i}{r_0}\right)^{\xi}}},
\end{equation}
which is independent of $\xi$ in this asymptotic regime.

\subsubsection{Inward geodesics ($v_i<0$) and reversal points}

For inward geodesics, $v_i<0$ and the radial coordinate is not restricted to $r>r_i$. In this case, the radial velocity vanishes not only at the throat $r=r_0$, but also at an additional radius given by
\begin{equation}
    r_{zero}=\left (\frac{1}{r_i^2}+\frac{v_{i}^2}{L^2\left(1-\left(\frac{r_i}{r_0}\right)^\xi\right)} \right )^{-\frac{1}{2}},
    \label{eqn:rzero}
\end{equation}
which defines a second turning point of the motion.

To determine whether these points correspond to circular orbits, we compute the radial acceleration,
\begin{multline}
   \ddot{r}=\frac{\xi}{2r}\left(\frac{r}{r_0}\right)^{\xi}\left[L^2\left(\frac{1}{r^2}- \frac{1}{r_i^2}\right)+\frac{v_{i}^2}{1-\left(\frac{r_i}{r_0}\right)^{\xi}}\right]\\
   +\frac{L^2}{r^3}\left(1-\left(\frac{r}{r_0}\right)^{\xi}\right).
\end{multline}

For circular orbits, one requires both $\dot{r}=0$ and $\ddot{r}=0$. Although $\dot{r}=0$ at $r=r_0$ and $r=r_{zero}$, the radial acceleration does not vanish at these points. Therefore, no circular orbits exist within this family of trajectories.

\begin{figure}[h]

\centering
\includegraphics[width=0.5\textwidth]{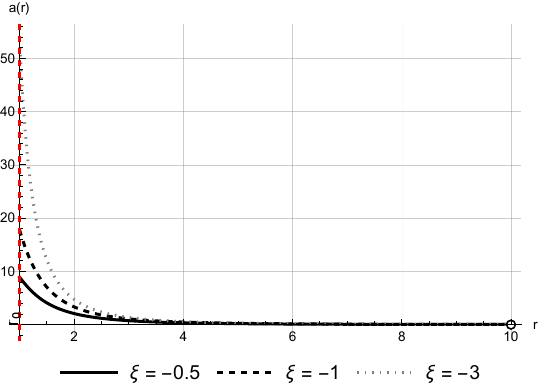}
\includegraphics[width=0.5\textwidth]{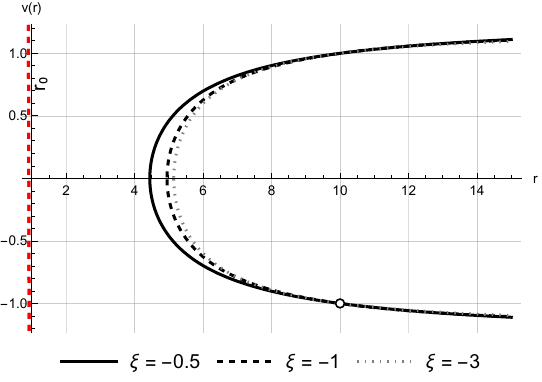}
\caption{
Top panel: radial acceleration for zero initial velocity, showing that $\ddot{r}>0$ for all $r>r_0$, so that any particle reaching zero radial velocity is driven outward. Bottom panel: trajectory of a particle with initial position $r_i$ and negative velocity $v_i$, illustrating the existence of a reversal point $r_{\mathrm{zero}}$ where the particle is deflected before reaching the throat. The vertical red dashed line in each panel indicates the wormhole throat at \(r=r_0\).
}
\label{InwardL}
\end{figure}

Instead, these radii correspond to \textit{reversal points} of the trajectory. In particular, since $\ddot{r}>0$ at $\dot{r}=0$, an infalling particle is deflected and reverses its motion before reaching the throat whenever $r_{zero}>r_0$ (see Fig.~\ref{InwardL}).

The condition for the existence of such reversal points outside the throat is
\begin{equation}
    r_{zero}=\frac{L}{\sqrt{E^2-1}} \geq r_0.
\end{equation}

Notably, this condition is independent of the parameter $\xi$ and depends only on the conserved quantities $E$ and $L$. Nevertheless, the geometry still enters implicitly through the definition of $E$. 

\subsubsection{Traversability of the wormhole}

It is convenient to describe the geometry in terms of the proper radial distance $l(r)$, defined by
\begin{equation}
    dl=\pm \frac{dr}{\sqrt{1-\left(\frac{r}{r_0}\right)^\xi}}.
\end{equation}

Fixing the throat at $l(r_0)=0$, one finds that the proper distance coincides with the embedding function,
\begin{equation}
    l(r)=\pm z(r),
    \label{eqn:properdistance}
\end{equation}
so that the wormhole connects two asymptotic regions with $-\infty<l<\infty$.

In terms of $l$, the radial equation of motion \eqref{eqn:geodesic} takes the form
\begin{equation}\label{dot(l)}
    \dot{l}^2=E^2-V(l)^2,
\end{equation}
where
\begin{equation}
    V(l)=\sqrt{1+\frac{L^2}{r(l)^2}}
\end{equation}
plays the role of an effective potential.

Within this description, traversability reduces to a simple condition on the conserved energy,
\begin{equation}
E^2\geq V(0)^2 = 1+\frac{L^2}{r_0^2},
\end{equation}
which ensures that the particle can cross the throat.

Conversely, if $E^2<V(0)^2$, the motion is bounded and the particle cannot reach the throat, instead reversing its trajectory at a finite proper distance. The corresponding turning point can be expressed in terms of the proper coordinate by evaluating \eqref{eqn:properdistance} at $r=r_{zero}$, yielding
\begin{equation}
    l_{zero} = \pm\frac{L}{\sqrt{E^2-1}}\,
    {}_{2}F_1\left( \frac{1}{2}, \frac{1}{\xi}, 1 + \frac{1}{\xi}, \left( \frac{L}{r_0\sqrt{E^2-1}} \right)^\xi \right).
\end{equation}

\subsubsection{Orbits}

To analyze the orbital structure, it is useful to rewrite the geodesic equations in terms of the inverse radial coordinate $u=1/r$. From the conserved angular momentum,
\begin{equation}
    \dot{\phi}=\frac{L}{r^2},
\end{equation}
it follows that
\begin{equation}
    L\dot{\phi}=\frac{L^2}{r^2}.
\end{equation}
Using Eq.~\eqref{eqn:geodesic}, this can be rewritten as
\begin{equation}
    L\dot{\phi}=E^2-1-\frac{\dot{r}^2}{1 - \left( \dfrac{r}{r_0} \right)^\xi}.
\end{equation}

Now, defining $u=1/r$, we have
\begin{eqnarray}
    \dot{\phi}=Lu^2, \\
    \dot{r}=-L\frac{du}{d\phi}.
\end{eqnarray}

If we further define $u_0=1/r_0$, then the orbital equation becomes
\begin{equation}
    u^2=\frac{E^2-1}{L^2}-\frac{\left (\frac{du}{d\phi}\right )^2}{1 - \left( \dfrac{u_0}{u} \right)^\xi}.
\end{equation}

Introducing
\begin{equation}
   u_{zero}^2=\frac{E^2-1}{L^2},
\end{equation}
we can rewrite this as
\begin{equation}
   \left (1 - \left( \dfrac{u_0}{u} \right)^\xi \right )\left ( u_{zero} ^2-u^2\right )=\left (\frac{du}{d\phi}\right )^2.
\end{equation}

Rearranging terms, we obtain
\begin{equation}
   \phi(u)=\pm \int \frac{du}{\sqrt{ \left (1 - \left( \dfrac{u_0}{u} \right)^\xi \right )\left ( u_{zero}^2-u^2\right )}}+C.
\end{equation}

Returning to the radial coordinate, we define
\begin{equation}
    \mathcal{F}(r)=r ^2\sqrt{\left (1 - \left( \dfrac{r}{r_0} \right)^\xi \right )\left (
   \dfrac{1}{r_{zero}^2}-\dfrac{1}{r ^2}\right )},
\end{equation}
so that, imposing $\phi(r_i)=0$, the orbit equation becomes
\begin{equation}
   \phi(r)=\pm\int_{r_i}^r\frac{dr}{\mathcal{F}(r)}.
   \label{eqn:orbit}
\end{equation}

As we are interested in infalling trajectories, we focus on solutions with $\dot r<0$, for which $du/d\phi>0$ and the negative branch in Eq.~\eqref{eqn:orbit} must be chosen. Two qualitatively different cases arise.

If $r_{zero}>r_0$, the particle reaches the turning point $r=r_{zero}$ before the throat and is reflected. In this case,
\begin{equation}
   \phi(r)=
   \begin{cases}
   -\int_{r_i}^r\frac{dr}{\mathcal{F}(r)}, & \dot{r}\leq 0, \\[3mm]
   -\int_{r_i}^{r_{zero}}\frac{dr}{\mathcal{F}(r)}
   +\int_{r_{zero}}^{r}\frac{dr}{\mathcal{F}(r)}, & \dot{r}>0.
   \end{cases}
\end{equation}

On the other hand, if $r_{zero}<r_0$, the particle crosses the throat and the radial velocity changes sign there. The orbit is then given by
\begin{equation}
   \phi(r)=
   \begin{cases}
   -\int_{r_i}^r\frac{dr}{\mathcal{F}(r)}, & \dot{r}\leq 0, \\[3mm]
   -\int_{r_i}^{r_0}\frac{dr}{\mathcal{F}(r)}
   +\int_{r_0}^{r}\frac{dr}{\mathcal{F}(r)}, & \dot{r}>0.
   \end{cases}
\end{equation}

Both types of orbits are shown in Fig.~\ref{AllOrbits} for different values of $\xi$.

\bibliography{thisbib}

@article{Sotiriou_2010,
  title = {$f(R)$ theories of gravity},
  author = {Sotiriou, Thomas P. and Faraoni, Valerio},
  journal = {Rev. Mod. Phys.},
  volume = {82},
  issue = {1},
  pages = {451--497},
  numpages = {0},
  year = {2010},
  month = {Mar},
  publisher = {American Physical Society},
  doi = {10.1103/RevModPhys.82.451},
  url = {https://link.aps.org/doi/10.1103/RevModPhys.82.451}
}

@article{Agrawal_2023,
   title={Unimodular gravity traversable wormholes},
   volume={138},
   ISSN={2190-5444},
   url={http://dx.doi.org/10.1140/epjp/s13360-023-03872-y},
   DOI={10.1140/epjp/s13360-023-03872-y},
   number={3},
   journal={The European Physical Journal Plus},
   publisher={Springer Science and Business Media LLC},
   author={Agrawal, A. S. and Mishra, B. and Moraes, P. H. R. S.},
   year={2023},
   month=mar }

@article{MorrisThorne1988,
  author = {Morris, Michael S. and Thorne, Kip S.},
  title = {Wormholes in spacetime and their use for interstellar travel: A tool for teaching general relativity},
  journal = {American Journal of Physics},
  volume = {56},
  number = {5},
  pages = {395--412},
  year = {1988},
  doi = {10.1119/1.15620}
}

@article{Visser1989,
   title={Traversable wormholes: Some simple examples},
   volume={39},
   ISSN={0556-2821},
   url={http://dx.doi.org/10.1103/PhysRevD.39.3182},
   DOI={10.1103/physrevd.39.3182},
   number={10},
   journal={Physical Review D},
   publisher={American Physical Society (APS)},
   author={Visser, Matt},
   year={1989},
   month=may, pages={3182–3184} }

@article{MorrisThorneYurtsever1988,
  author = {Morris, Michael S. and Thorne, Kip S. and Yurtsever, Ulvi},
  title = {Wormholes, time machines, and the weak energy condition},
  journal = {Physical Review Letters},
  volume = {61},
  number = {13},
  pages = {1446--1449},
  year = {1988},
  doi = {10.1103/PhysRevLett.61.1446}
}

@article{PastenBosquezCruz2026,
    author = "Past{\'e}n, Erick and Bosquez, Marco and Cruz, Norman",
    title = "{Null Raychaudhuri equation and the impossibility of traversable wormholes in unimodular gravity}",
    eprint = "2602.00524",
    archivePrefix = "arXiv",
    primaryClass = "gr-qc",
    doi = "10.1088/1361-6382/ae73df",
    journal = "Class. Quant. Grav.",
    volume = "43",
    number = "11",
    pages = "115014",
    year = "2026"
}

@article{CataldoCruz2025,
  author = {Cataldo, Mauricio and Cruz, Norman},
  title = {Revisiting wormhole solutions in unimodular gravity: Energy conditions and exotic matter requirements},
  journal = {The European Physical Journal Plus},
  volume = {140},
  pages = {1014},
  year = {2025},
  doi = {10.1140/epjp/s13360-025-06960-3}
}

@INCOLLECTION{Einstein1919,
       author = {{Einstein}, A.},
        title = "{Do gravitational fields play an essential part in the structure of the elementary particles of matter?}",
    booktitle = {The Principle of Relativity. Dover Books on Physics. June 1},
         year = 1952,
        pages = {189-198},
       adsurl = {https://ui.adsabs.harvard.edu/abs/1952prel.book..189E}
}

@article{AndersonFinkelstein1971,
    author = "Anderson, J. L. and Finkelstein, D.",
    title = "{Cosmological constant and fundamental length}",
    doi = "10.1119/1.1986321",
    journal = "Am. J. Phys.",
    volume = "39",
    pages = "901--904",
    year = "1971"
}

@article{Unruh1989,
    author = "Unruh, W. G.",
    title = "{A Unimodular Theory of Canonical Quantum Gravity}",
    reportNumber = "NSF-ITP-88-169",
    doi = "10.1103/PhysRevD.40.1048",
    journal = "Phys. Rev. D",
    volume = "40",
    pages = "1048",
    year = "1989"
}

@article{Alvarez2015,
   title={Unimodular gravity redux},
   volume={92},
   ISSN={1550-2368},
   url={http://dx.doi.org/10.1103/PhysRevD.92.061502},
   DOI={10.1103/physrevd.92.061502},
   number={6},
   journal={Physical Review D},
   publisher={American Physical Society (APS)},
   author={Álvarez, E. and González-Martín, S. and Herrero-Valea, M. and Martín, C.P.},
   year={2015},
   month=sep }

@article{NojiriOdintsov2011,
   title={Unified cosmic history in modified gravity: From F(R) theory to Lorentz non-invariant models},
   volume={505},
   ISSN={0370-1573},
   url={http://dx.doi.org/10.1016/j.physrep.2011.04.001},
   DOI={10.1016/j.physrep.2011.04.001},
   number={2–4},
   journal={Physics Reports},
   publisher={Elsevier BV},
   author={Nojiri, Shin’ichi and Odintsov, Sergei D.},
   year={2011},
   month=aug, pages={59–144} }

@article{Weinberg1989,
    author = "Weinberg, Steven",
    editor = "Hsu, Jong-Ping and Fine, D.",
    title = "{The Cosmological Constant Problem}",
    reportNumber = "UTTG-12-88",
    doi = "10.1103/RevModPhys.61.1",
    journal = "Rev. Mod. Phys.",
    volume = "61",
    pages = "1--23",
    year = "1989"
}

@article{Visser2003,
  author  = {Visser, Matt and Kar, Sayan and Dadhich, Naresh},
  title   = {Traversable wormholes with arbitrarily small energy condition violations},
  journal = {Phys. Rev. Lett.},
  volume  = {90},
  pages   = {201102},
  year    = {2003},
  doi     = {10.1103/PhysRevLett.90.201102}
}

@article{Lobo2005,
  author  = {Lobo, Francisco S. N.},
  title   = {Energy conditions, traversable wormholes and dust shells},
  journal = {Gen. Rel. Grav.},
  volume  = {37},
  pages   = {2023},
  year    = {2005},
  doi     = {10.1007/s10714-005-0181-9}
}

@article{Einstein1915,
  author  = {Einstein, Albert},
  title   = {Die Feldgleichungen der Gravitation},
  journal = {Sitzungsberichte der Königlich Preussischen Akademie der Wissenschaften zu Berlin},
  year    = {1915},
  pages   = {844--847}
}

@article{Fabris_2022,
   title={Nonconservative unimodular gravity: a viable cosmological scenario?},
   volume={82},
   ISSN={1434-6052},
   url={http://dx.doi.org/10.1140/epjc/s10052-022-10470-2},
   DOI={10.1140/epjc/s10052-022-10470-2},
   number={6},
   journal={The European Physical Journal C},
   publisher={Springer Science and Business Media LLC},
   author={Fabris, Júlio C. and Alvarenga, Marcelo H. and Hamani-Daouda, Mahamadou and Velten, Hermano},
   year={2022},
   month=June }

@article{Ellis_2011,
doi = {10.1088/0264-9381/28/22/225007},
url = {https://doi.org/10.1088/0264-9381/28/22/225007},
year = {2011},
month = {oct},
publisher = {IOP Publishing},
volume = {28},
number = {22},
pages = {225007},
author = {Ellis, George F R and van Elst, Henk and Murugan, Jeff and Uzan, Jean-Philippe},
title = {On the trace-free Einstein equations as a viable alternative to general relativity},
journal = {Classical and Quantum Gravity}
}
\end{document}